\documentclass[12pt]{iopart}
\usepackage{graphicx}
\usepackage{iopams}
\begin{document}
\bibliographystyle{unsrt}
\title[Magnetic susceptibility of YbRh$_{2}$Si$_{2}$ and YbIr$_{2}$Si$_{2}$]{Magnetic susceptibility of YbRh$_{2}$Si$_{2}$ and YbIr$_{2}$Si$_{2}$ on the basis of a localized 4$f$ electron approach}

\author{A S Kutuzov$^1$, A M Skvortsova$^1$, S I Belov$^1$, \\
             J Sichelschmidt$^2$, J Wykhoff$^2$, I Eremin$^3$, C Krellner$^2$, \\
             C Geibel$^2$ and B I Kochelaev$^1$}
\address{$^1$Physics Department, Kazan State University, 420008 Kazan, Russia\\
			$^2$Max Planck Institute for Chemical Physics of Solids, 01187 Dresden, Germany\\
			$^3$Max Planck Institute for the Physics of Complex Systems, 01187 Dresden, Germany}
\ead{Sichelschmidt@cpfs.mpg.de}
\date{\today}

\begin{abstract} 	
We consider the local properties of the Yb$^{3+}$ ion in the crystal electric field in the Kondo lattice compounds YbRh$_{2}$Si$_{2}$ and YbIr$_{2}$Si$_{2}$. On this basis we have calculated the magnetic susceptibility taking into account the Kondo interaction in the simplest molecular field approximation. The resulting Curie-Weiss law and Van Vleck susceptibilities could be excellently fitted to experimental results in a wide temperature interval where thermodynamic and transport properties show non-Fermi-liquid behaviour for these materials.
\end{abstract}

\pacs{75.40.Cx, 75.30.Mb, 76.30.Kg, 75.10.Dg}

\maketitle
%
\section{Introduction}
Very peculiar magnetic, thermal, and transport properties of 4$f$ electron based heavy fermion systems are determined by the interplay of the strong repulsion of 4$f$ electrons on the rare-earth ion sites, their hybridization with wide band conduction electrons, and an influence of the crystalline electrical field. Consequences of the mentioned interplay for the electronic energy band structure near the Fermi-energy ($E_F$) were recently studied in YbRh$_2$Si$_2$ and YbIr$_2$Si$_2$ by angle-resolved photoemission and interpreted within the periodic Anderson model \cite{danzenbacher07a, vyalikh08a}. It was found that the hybridization of 4$f$ electrons results in a rather flat 4$f$ band near $E_F$ . Additionally, renormalization of the valence state leads to the formation of a heavy band that reveals strong 4f character close to $E_F$. Moreover, slow valence fluctuations of the Yb ion may occur between $4f^{13}$ and closed $4f^{14}$ configurations with an averaged valence value of about +2.9 \cite{knebel06a}. Evidently, these observations are consistent with a metallic behavior with very heavy charge carriers having properties of a Landau Fermi-liquid (LFL). At the same time the thermal, magnetic, and transport measurements show that the heavy fermions with a well defined Fermi-surface survive only at very low temperatures, coexisting with long-range antiferromagnetic (AF) order ($T_N = 70$~mK in YbRh$_2$Si$_2$) which is suppressed at a magnetic quantum critical point (QCP) \cite{gegenwart08a} by an external magnetic field (at $H_c = 600$~G, $T_N \rightarrow 0$~K if c axis $\perp H$ ). With further increasing magnetic field at temperatures below a characteristic temperature $T^*$, roughly proportional to $H-H_c$ , a crossover to LFL behavior is found. At temperatures above $T_N$ and $T^*$, but below the single ion Kondo temperature $T_K$, the properties of the discussed materials are quite unusual and display a non-Fermi liquid (NFL) behavior. The underlying fluctuations at the QCP are discussed to be locally critical, i.e. all the low-energy degrees of freedom have an atomic length scale \cite{gegenwart08a, si01a}. One of the hallmarks of this local criticality is a generalized Curie-Weiss law like $\chi \propto T^{-\alpha} + const$ for the magnetic susceptibility with an exponent $\alpha < 1$ \cite{si01a, schroder00a}. This type of behavior with $\alpha = 0.75$ was found first in CeCu$_{5.9}$Au$_{0.1}$ \cite{schroder00a}. The peculiarities of these properties are related to the competition between two interactions, originating both from the above mentioned hybridization: an exchange coupling of the local moments with the broadband conduction electrons (Kondo interaction) and an induced indirect RKKY interaction between the moments.
	The importance of the local properties in magnetic dynamics was mainly confirmed by the discovery of a strong and rather narrow electron paramagnetic resonance (EPR) in YbRh$_2$Si$_2$ and YbIr$_2$Si$_2$ below the Kondo temperature $T_K$ = 25 K and $T_K$ = 40 K, respectively ($T_{\rm K}$ revealed by specific heat data) \cite{sichelschmidt03a, sichelschmidt07a, hossain05a}. This EPR signal was quite unexpected, since it was believed that the Yb$^{3+}$ magnetic moment at $T<T_K$ should be screened by the conduction electrons and that the EPR linewidth should reach large values $\Delta H \propto k_ B T_K / g\mu_B$ by approaching to $T_K$ from above. Moreover, the main features of the observed EPR signal (anisotropy of the g factor and the EPR linewidth) reflect local properties of the Yb$^{3+}$ ion in the crystal electric field. The integrated intensity of the EPR line is proportional to the homogeneous static magnetic susceptibility.
Having these experimentally confirmed local properties at hand we assume entirely local properties of the Yb$^{3+}$ ion in the crystal electric field in order to theoretically investigate the energy spectra, g factors of the ground state, and the static magnetic susceptibility of YbRh$_2$Si$_2$ and YbIr$_2$Si$_2$ compounds.
%
%
\section{Yb$^{3+}$ ion in tetragonal crystal field}
A free Yb$^{3+}$ ion has a $4f^{13}$ configuration with one term $^2F$. As the spin-orbital coupling is much stronger than the crystal field in our compounds, the total momentum $\mathbf{J}$ is a good quantum number. The spin-orbital interaction splits the $^2F$ term into two multiplets: $^2F_{7/2}$ with total momentum $J = 7/2$ and $^2F_{5/2}$ with $J = 5/2$. Both are separated by about 1 eV \cite{abragam70a} and therefore we will consider only the ground multiplet $^2F_{7/2}$. The potential of the tetragonal crystal field for an ion can be written as
\begin{equation}
V=\alpha B^0_2 O^0_2 + \beta (B^0_4 O^0_4 + B^4_4 O^4_4) + \gamma (B^0_6 O^0_6 + B^4_6 O^4_6).
\label{eq1}
\end{equation}
To define energy levels and wave functions of the Yb$^{3+}$ ion we have to diagonalize the matrix of the operator (\ref{eq1}) on the states of the ground multiplet $^2F_{7/2}$. In (\ref{eq1}) $B^q_k$ are crystal field parameters, the operators $O^q_k(\mathbf{J})$ their matrix elements and $\alpha = 2/63, \beta = -2/1155, \gamma = 4/27027$ are given in \cite{abragam70a}.\\
As follows from the group theory, the two-valued irreducible representation $D^{7/2}$ of rotation group contains two two-dimensional irreducible representations $\Gamma^t_7$ and $\Gamma^t_6$ of the double tetragonal group $D^{7/2}=2\Gamma^t_7 + 2\Gamma^t_6$  \cite{abragam70a}. Therefore the states of Yb$^{3+}$ in a tetragonal field are four Kramers doublets and to diagonalize Hamiltonian (\ref{eq1}) we just need to diagonalize two two-dimensional matrices corresponding to the representations $\Gamma^t_7$ and $\Gamma^t_6$. Hence, the crystal field splits the lower $^2F_{7/2}$ multiplet into four Kramers doublets with energies, wave functions and $g$ factors as given in Table 1.
\Table{\label{tab1}Energies, wave functions and $g$ factors of Yb$^{3+}$ ion in a tetragonal crystal field for  $\Gamma^t_7$ and $\Gamma^t_6$ representations.}
\br
$E_{1,2}(\Gamma^t_7)=-D\pm C/\cos\varphi_c$ & $E_{3,4}(\Gamma^t_6)=D\pm A/\cos\varphi_a$ \\
\mr
$|1\uparrow,2\uparrow\rangle = \pm c_{1,2}|5/2\rangle+c_{2,1}|-3/2\rangle$ & $|3\uparrow,4\uparrow\rangle = \pm a_{1,2}|-7/2\rangle+a_{2,1}|1/2\rangle$\\
$|1\downarrow,2\downarrow\rangle = \mp c_{1,2}|-5/2\rangle-c_{2,1}|3/2\rangle$ & $|3\downarrow,4\downarrow\rangle = \mp a_{1,2}|7/2\rangle-a_{2,1}|-1/2\rangle$\\
\mr
$g^{1,2}_\|(\Gamma^t_7)=g_J(1\pm 4\cos\varphi_c)$ & $g^{3,4}_\|(\Gamma^t_6)=-g_J(3\pm 4\cos\varphi_a)$ \\
$g^{1,2}_\perp(\Gamma^t_7)=\mp 2\sqrt{3}g_J\sin\varphi_c$ & $g^{3,4}_\perp(\Gamma^t_6)=-2g_J(1\mp\cos\varphi_a)$ \\
\br
\end{tabular}
\end{indented}
\end{table}
In this table upper and lower signs correspond to left and right indexes; $\uparrow,\downarrow$ correspond to the Kramers doublets effective spin projection up and down, $c_1 = \cos(\varphi_c /2)$, $c_2 = \sin(\varphi_c /2)$, $a_1 = \cos(\varphi_a /2)$, $a_2 = \sin(\varphi_a /2)$, $\tan\varphi_a = \tilde{A}/A$, $\tan\varphi_c = \tilde{C}/C$, $-\pi/2\le\varphi_a$, $\varphi_c\le\pi/2$, and $g_J = 8/7$ is the Land\'{e} $g$ factor. We used parameters $A, C, D, \tilde{A}, \tilde{C}$ which are defined by the crystal field parameters:
\begin{eqnarray}
\fl A= 4B^0_2/7 + 8B^0_4/77 + 80B^0_6/143, \;C=4B^0_2/21 + 40B^0_4/77-560B^0_6/429,  \nonumber\\ 
\fl D=2B^0_2/21 - 64B^0_4/77 - 160B^0_6/429, \\
\fl \tilde{A}= -8\sqrt{35}B^4_4/385 + 80\sqrt{35}B^4_6/3003, \; \tilde{C}= -8\sqrt{3}B^4_4/77 - 80\sqrt{3}B^4_6/1287.  \nonumber 
\end{eqnarray}
The Zeeman energy $g_J\mu_B\mathbf{HJ}$ in the basis $|m\sigma\rangle \,(m = 1...4, \sigma = \uparrow, \downarrow)$ of each doublet could be represented by
\begin{equation}
\mathcal{H}_{\rm Zeeman}=g_\|\mu_BH_zS_z + g_\perp\mu_B(H_xS_x + H_yS_y)
\label{eq3}
\end{equation}
where $\mathbf{H}$ is the magnetic field, $\mathbf{ S}$ is the effective spin operator with $S=1/2$, $\mu_B$ is the Bohr magneton, $g_\|$ and $g_\perp$ are $g$ factors when the field is applied parallel and perpendicular to the c-axis, respectively (Table 1).\\
As was mentioned above the main features of EPR signal observed in YbRh$_{2}$Si$_{2}$ and YbIr$_{2}$Si$_{2}$ reflect the local properties of Yb$^{3+}$ ion. The EPR signal in YbRh$_{2}$Si$_{2}$ and YbIr$_{2}$Si$_{2}$ is highly anisotropic \cite{sichelschmidt07a, sichelschmidt07b}. The angular dependence of $g$ factors in these compounds is well described by
\begin{equation}
g=\sqrt{g^2_\|\cos^2\theta + g^2_\perp\sin^2\theta},
\end{equation}
where $\theta$ is the angle between magnetic field and crystal c-axis orientations, with $|g_\|| = 0.17$, $|g_\perp | = 3.56$ for YbRh$_{2}$Si$_{2}$ and $|g_\|| = 0.855$, $|g_\perp | = 3.36$ for YbIr$_{2}$Si$_{2}$ at $T = 5$~K. From neutron scattering experiments \cite{stockert06a, hiess06a} the intervals between the ground Kramers doublet and the exited energy levels amount to $\Delta_1 = 17$~meV, $\Delta_2 = 25$~meV, $\Delta_3 = 43$~meV for YbRh$_{2}$Si$_{2}$ and $\Delta_1 = 18$~meV, $\Delta_2 = 25$~meV, $\Delta_3 = 36$~meV for YbIr$_{2}$Si$_{2}$. Unfortunately, these four independent values (three energy intervals and one parameter which define $g_\|$ and $g_\perp$, see Table 1) do not allow to determine five crystal field parameters unambiguously.\\
Figure 1 represents the diagram of $g$ factors together with the experimental points for the effective $g$ factors of YbRh$_{2}$Si$_{2}$ and YbIr$_{2}$Si$_{2}$ (four points with different signs of $g_\|$ and $g_\perp$). The solid and dashed parts of the line $g_\| + 2g_\perp + 7g_J=0$  in Fig. 1 correspond to the doublets $E_4(\Gamma^t_6)$ and $E_3(\Gamma^t_6)$, and the solid and dashed parts of the ellipse $(g_\| -g_J)^2/4+ g_\perp^2/3=4g_J^2$ correspond to the doublets $E_2(\Gamma^t_7)$ and $E_1(\Gamma^t_7)$. It is evident from the proximity to the data (symbols in Fig. 1) that only the doublets $E_2(\Gamma^t_7)$ and $E_4(\Gamma^t_6)$ could be considered for the ground state.
\begin{figure}
 \centering
 \includegraphics{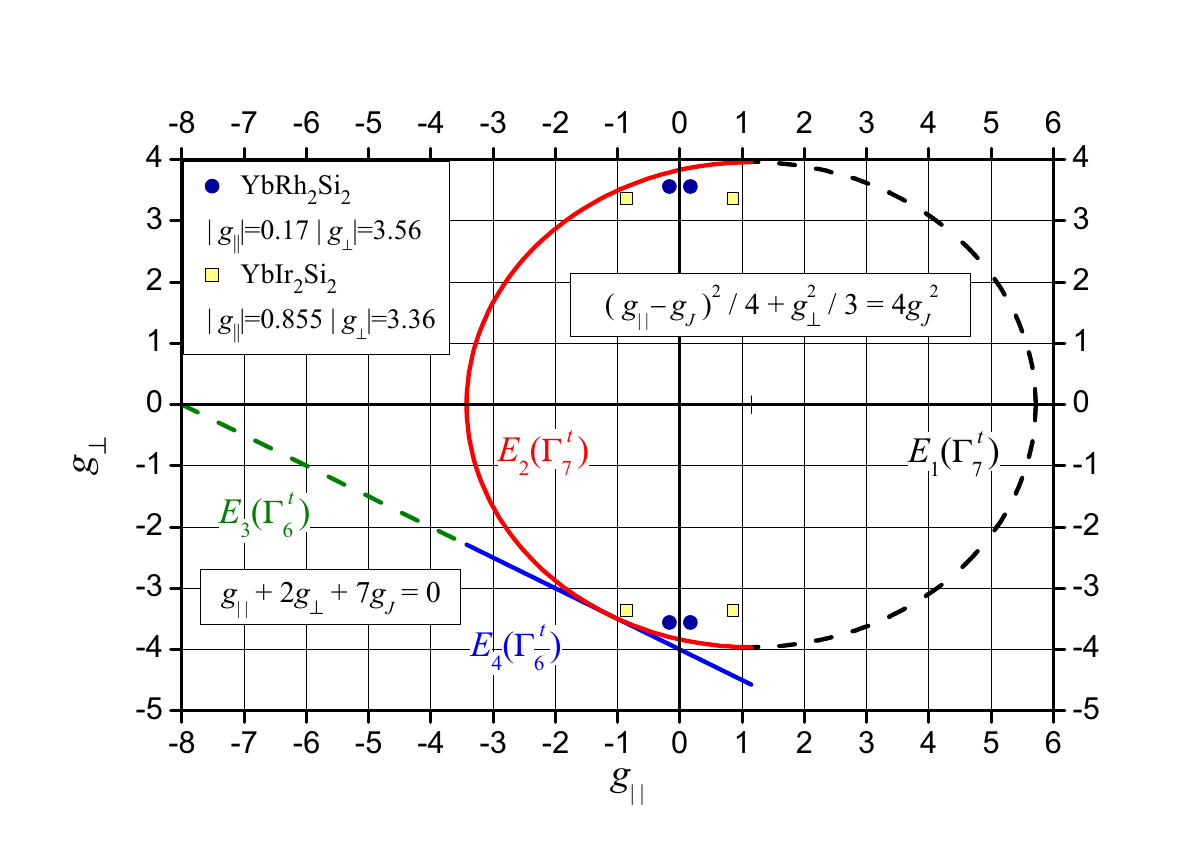}
\caption{Diagram of $g$ factors and experimental points at $T=5$~K for effective $g$ factors of YbRh$_{2}$Si$_{2}$ and YbIr$_{2}$Si$_{2}$. \label{Fig1}}
\end{figure}
The theoretical $g$ values (i.e. the points on the ellipse and on the line nearest to the experimental points) and the corresponding values of $\varphi_c$ and $\varphi_a$ are given in Table 2. These results are qualitatively well consistent with the experimental ones being at the same time somewhat larger than the measured ones. Similar
\Table{\label{tab2}Optimal theoretical $g$ factors of Yb$^{3+}$ ion in YbRh$_{2}$Si$_{2}$ and YbIr$_{2}$Si$_{2}$.}
\br
Compound                                                        & Ground doublet, $\varphi_c$, $\varphi_a$ & $g_\|$ & $g_\perp$ \\ 
\mr
                                                                          & $E_2(\Gamma^t_7), \varphi_c=\pm1.2660, -\pi/2\le\varphi_a\le\pi/2$ & -0.229 & $\pm3.777$ \\  
\raisebox{1.5ex}[-1.5ex]{YbRh$_{2}$Si$_{2}$} & $E_4(\Gamma^t_6), \varphi_a=\pm0.8206, -\pi/2\le\varphi_c\le\pi/2$ & -0.312 & -3.844 \\ \mr
                                                                          & $E_2(\Gamma^t_7), \varphi_c=\pm1.1003, -\pi/2\le\varphi_a\le\pi/2$ & -0.929 & $\pm3.529$ \\ 
\raisebox{1.5ex}[-1.5ex]{YbIr$_{2}$Si$_{2}$}   & $E_4(\Gamma^t_6), \varphi_a=\pm0.9952, -\pi/2\le\varphi_c\le\pi/2$ & -0.940 & -3.530 \\ \br
\end{tabular}
\end{indented}
\end{table}
estimations for $g$ factors were obtained recently in \cite{leushin07a}. A slight difference between experimental and our theoretical values can be explained mainly by taking into account the above mentioned the Kondo interaction, i.e. an exchange coupling between the 4$f$ electrons of the Yb$^{3+}$ ion and wide-band conduction electrons. This interaction becomes highly anisotropic after projection onto the ground Kramers doublet:
\begin{equation}
\mathcal{H}_{\rm int}=-\sum_i\Big\{J^\perp_{s\sigma}\left[S^x_i\sigma^x(\mathbf{r}_i) + S^y_i\sigma^y(\mathbf{r}_i)\right] + J^\|_{s\sigma}S^z_i\sigma^z(\mathbf{r}_i)\Big\}.
\label{eq5}
\end{equation}
Here $\sigma(\mathbf{r}_i)$ is the operator of the conduction electrons spin density, and $J^{\|,\perp}$ are the exchange coupling integrals. This leads to the so-called Knight shift of the $g$ factor. In the case of the anisotropic exchange interaction of the antiferromagnetic sign ($J^{\|,\perp}<0$) this shift reduces the absolute value of the ionic $g$ factor the same way as it happens in the isotropic case \cite{barnes81a, kochelaev04a}:
\begin{equation}
g^{\rm eff}_{\|,\perp}=g^0_{\|,\perp}(1+\lambda_{\|,\perp}\chi_\sigma)\,, \; \lambda_{\|,\perp}=\frac{J^{\|,\perp}_{s\sigma}}{g_{\|,\perp}g_\sigma\mu^2_B}
\end{equation}
$g^0_{\|,\perp}$ is the ionic $g$ factor, $g_\sigma$  is the $g$ factor of conduction electrons, $\chi_\sigma$ is the Pauli magnetic susceptibility, and $\lambda_{\|,\perp}$ are molecular field constants. We also have performed an improvement of this simplest contribution of the Kondo interaction by the methods of a renormalization group analysis. Then, as will be published elsewhere, critical terms like $\ln^{-1}(T/T^{\|,\perp}_K)$ appear which reduce further the $g$ factor values at low temperatures.
\section{Static magnetic susceptibility}
The magnetization of the crystal is $n\langle M\rangle$, where $n$ is the ion concentration, $M=-g_J\mu_B\mathbf{J}$ is the ionÕs magnetic moment operator, $\langle M_\alpha \rangle = \Tr(\exp(-\beta \mathcal{H})M_\alpha) / \Tr \exp(-\beta \mathcal{H})$ is the mean value of the $\alpha$ component of magnetic moment, $\beta = 1/(k_BT)$, $k_B$ is the Boltzmann constant, and $T$ is the temperature. $\mathcal{H} = V-\mathbf{MH}$ is the Hamiltonian, where $V$ is the crystal field potential (\ref{eq1}) of the Yb$^{3+}$ ion, and the second term corresponds to the interaction of the magnetic moment with the magnetic field $\mathbf{H}$.\\
The magnetic susceptibility is defined as
\begin{equation}
\label{eq7}
\chi_{\alpha\gamma}=n\left.\frac{\partial\langle M_\alpha\rangle}{\partial H_\gamma}\right |_{\mathbf{H}=0}=n\int^\beta_0\langle M_\alpha M_\gamma(\lambda)\rangle_0d\lambda\,,
\end{equation}
where $\langle ...\rangle_0$ is calculated with $V, M_\alpha(\lambda)=\exp(-\lambda V)M_\alpha \exp(\lambda V)$. If we suppose the temperature to be such low that $\exp[-\beta(E_m-E_k)]\approx0$ for $m\neq k$ we distinguish two different contributions in the susceptibility $\chi=\chi^C+\chi^{VV}$,
\begin{eqnarray}
\label{eq8}
\chi^C_{\alpha\gamma}=\frac{n\beta(g_J\mu_B)^2}{2}\sum_{\sigma\sigma'}\langle k\sigma | J_\alpha |k\sigma'\rangle \langle k\sigma' | J_\gamma |k\sigma\rangle
                                       \equiv \frac{C^0_{\alpha\gamma}}{T}\,,\\
\chi^{VV}_{\alpha\gamma}=n(g_J\mu_B)^2\sum_{\stackrel{m(\neq k)}{\sigma\sigma '}} 
                                        \frac{\langle k\sigma | J_\alpha |m\sigma'\rangle \langle m\sigma' | J_\gamma |k\sigma\rangle}{E_m-E_k}\,.                                       
\end{eqnarray}
Here $\sigma, \sigma'=\uparrow, \downarrow$. The first term, $\chi^C_{\alpha\gamma}$, corresponds to the Curie susceptibility proportional to inverse temperature, the second term, $\chi^{VV}_{\alpha\gamma}$, corresponds to the Van Vleck susceptibility which does not depend on temperature. $|k\sigma\rangle$ indicates the states of the ground Kramers doublet. In the basis of the Kramers doublet states $|m\sigma\rangle$: $\chi_{xz}=\chi_{yz}=0$ and $\chi_{xx}=\chi_{yy}\equiv\chi_\perp$, which is the evident result for tetragonal symmetry. We also introduce $\chi_{zz}=\chi_\|$, and for the Curie constants $C^0_{zz}\equiv C^0_\|$, $C^0_{xx}=C^0_{yy}\equiv C^0_\perp$.\\
As was shown above, the Kramers doublets $E_2(\Gamma^t_7)$ and $E_4(\Gamma^t_6)$ describe the ground state properties almost equally well. For these two cases the Curie and Van Vleck parts of the susceptibility could be expressed by the parameters $c_i, a_i$ and energy intervals between Kramers doublets. If the ground state is $E_2(\Gamma^t_7)$ then
\begin{eqnarray}
\fl\chi^C_\|=n\beta(g_J\mu_B)^2\left ( \frac{5}{2}c^2_2-\frac{3}{2}c^2_1 \right)^2 \,,\; \chi^C_\perp=12n\beta(g_J\mu_B)^2c^2_1c^2_2\,,\nonumber \\ 
\fl\chi^{VV}_\|=\frac{32n(g_J\mu_B)^2c^2_1c^2_2}{E_1-E_2}\,,\\
\fl\chi^{VV}_\perp=n(g_J\mu_B)^2\left[\frac{6(c^2_1-c^2_2)^2}{E_1-E_2}+\frac{(\sqrt{7}c_2a_1-\sqrt{15}c_1a_2)^2}{2(E_3-E_2)}
                           +\frac{(\sqrt{7}c_2a_2+\sqrt{15}c_1a_1)^2}{2(E_4-E_2)}\right]. \nonumber \\ \nonumber
\end{eqnarray}
If the ground state is $E_4(\Gamma^t_6)$ then
\begin{eqnarray}
\fl\chi^C_\|=n\beta(g_J\mu_B)^2\left ( \frac{1}{2}a^2_1-\frac{7}{2}a^2_2 \right)^2 \,,\; \chi^C_\perp=4n\beta(g_J\mu_B)^2a^4_1\,,\nonumber \\ 
\fl\chi^{VV}_\|=\frac{32n(g_J\mu_B)^2a^2_1a^2_2}{E_3-E_4}\,,\\
\fl\chi^{VV}_\perp=n(g_J\mu_B)^2\left[\frac{8a^2_1a^2_2}{E_3-E_4}+\frac{(\sqrt{7}c_1a_2-\sqrt{15}c_2a_1)^2}{2(E_1-E_4)}
                           +\frac{(\sqrt{7}c_2a_2+\sqrt{15}c_1a_1)^2}{2(E_2-E_4)}\right]. \nonumber \\ \nonumber
\end{eqnarray}
The calculated molar susceptibility values are given in Table 3. Here we used the $g$ factors and parameters $\varphi_c$ and $\varphi_a$ from Table 2. In Table 3 the maximal and minimal possible values of $\chi^{VV}_{\|,\perp}$ for different excited doublets sequences and uncertain parameters $-\pi/2\le\varphi_a$, $\varphi_c\le\pi/2$ are given. We used for calculations experimental values of energy intervals $\Delta_i$.
\Table{\label{tab3}Calculated Curie constant $C^0$ $(10^{-6}\rm{m^3mol^{-1}K})$ and Van Vleck susceptibility $\chi^{VV}$ $(10^{-6}\rm{m^3mol^{-1}})$ for YbRh$_{2}$Si$_{2}$ and YbIr$_{2}$Si$_{2}$.}
\br
 & Ground  & & & & & \\
\ns \ns
Compound    &    & \centre{1}{$C^0_\perp$} & \centre{1}{$\chi^{VV}_\perp$} & \centre{1}{$C^0_\|$} & \centre{1}{$\chi^{VV}_\|$} \\ 
\ns \ns \ns
& doublet &  &  &  & \\ \mr                                                   
                                                                          & $E_2(\Gamma^t_7)$ & 16.8 & 0.087-0.202 & 0.062 & 0.09-0.227 \\ 
\raisebox{1.5ex}[-1.5ex]{YbRh$_{2}$Si$_{2}$} & $E_4(\Gamma^t_6)$ & 17.4 & 0.107-0.237 & 0.115 & 0.053-0.134 \\ \mr
                                                                          & $E_2(\Gamma^t_7)$ & 14.7 & 0.121-0.215 & 1.02 & 0.094-0.187 \\ 
\raisebox{1.5ex}[-1.5ex]{YbIr$_{2}$Si$_{2}$}   & $E_4(\Gamma^t_6)$  &  14.7 & 0.127-0.224 & 1.04 & 0.083-0.166 \\ \br
\end{tabular}
\end{indented}
\end{table}
It follows from our calculations that Curie-Weiss and Van Vleck susceptibilities play different roles in parallel and perpendicular orientations: for YbRh$_{2}$Si$_{2}$ the Curie constant in perpendicular orientation $C^0_\perp$ is at least two orders of magnitude larger than the Curie constant in parallel orientation $C^0_\|$ whereas the Van Vleck susceptibility $\chi^{VV}_\perp$ has the same order of magnitude as $\chi^{VV}_\|$. For YbIr$_{2}$Si$_{2}$ the situation is significantly different: $C^0_\perp$ and $C^0_\|$ differ only by one order of magnitude whereas the Van Vleck part is almost the same as for YbRh$_{2}$Si$_{2}$.
\section{Comparison with the experimental data }
It is evident that the measured susceptibility includes both the Yb$^{3+}$ ions and the conduction electrons susceptibilities. In the molecular field approximation the Kondo interaction and RKKY interactions renormalize the total susceptibility \cite{barnes81a, kochelaev04a, mattis88a}:
\begin{equation}
\label{eq12}
\chi_{\|,\perp}+\chi_\sigma=\frac{\chi^0_{\|,\perp}+\chi^0_\sigma+2\lambda_{\|,\perp}\chi^0_{\|,\perp}\chi^0_\sigma}{1-(\lambda^2_{\|,\perp}\chi^0_\sigma+\alpha_{\|,\perp})\chi^0_{\|,\perp}}\,,\;\; \chi^0_{\|,\perp}=\frac{C^0_{\|,\perp}}{T}\,,
\end{equation}
where $\alpha_{\|,\perp}$ are additional contributions to the molecular field from the RKKY interaction \cite{mattis88a}.\\
This renormalization leads to a Curie-Weiss law just as in the isotropic case \cite{barnes81a}. The Pauli susceptibility is negligible. We neglected also the renormalization of the Van Vleck part of the susceptibility. Finally in the molecular field approximation we can write for the total magnetic susceptibility:
\begin{equation}
\label{eq13}
\chi^{tot}_{\|,\perp}=\frac{C_{\|,\perp}}{T+\theta_{\|,\perp}} + \chi^{VV}_{\|,\perp}
\end{equation}
with
\begin{equation}
\label{eq14}
C_{\|,\perp}=C^0_{\|,\perp}(1+2\lambda_{\|,\perp}\chi^0_\sigma)
\end{equation}
and $\theta_{\|,\perp}$ independent on temperature. It is evident that we should expect $C<C^0$ if $\lambda<0$.\\
\begin{figure}[htbp]
  \centering
 \includegraphics{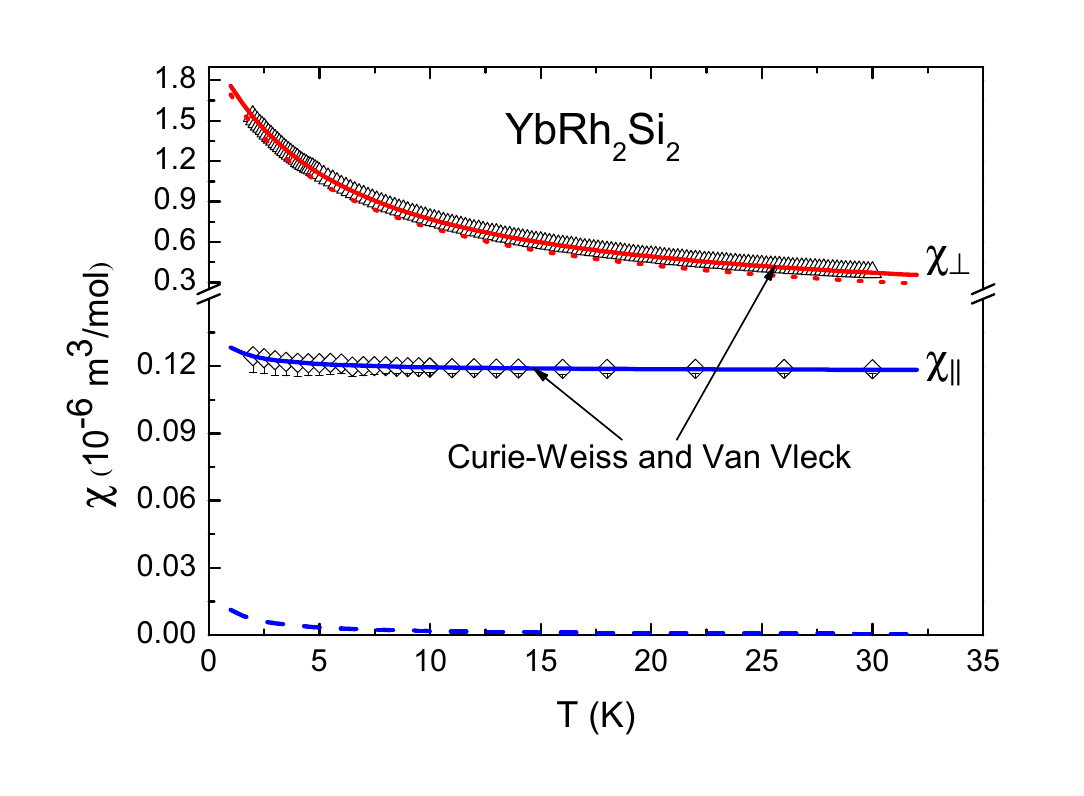}
  \caption{Static magnetic susceptibility of YbRh$_{2}$Si$_{2}$. Solid lines: fitting of susceptibility data with expression (\ref{eq13}) and parameters shown in Table 4. Dotted and dashed lines: contributions of Curie-Weiss part of susceptibility in perpendicular ($H = 11.2$~kG) and in parallel orientations ($H=10$~kG), correspondingly. Error bars indicate orientational precision. Note the different scales for $\chi_\perp$ and $\chi_\|$.\label{Fig2}}
 
\end{figure}
\noindent Figures 2 and 3 show the temperature dependence of susceptibility in YbRh$_{2}$Si$_{2}$ and YbIr$_{2}$Si$_{2}$ above 2K. It is easy to notice that the experimental data reflect the theoretically predicted tendencies according expression (\ref{eq13}). Indeed, for YbRh$_{2}$Si$_{2}$ in perpendicular orientation the main role is played by the Curie-Weiss contribution to susceptibility, but in parallel orientation the susceptibility is almost temperature independent and the main contribution comes from the Van Vleck part. For YbIr$_{2}$Si$_{2}$ the role of the Curie-Weiss susceptibility in parallel orientation is more important in comparison with YbRh$_{2}$Si$_{2}$.\\
Table 4 presents the values of the fitting parameters $C_\perp$, $\chi^{VV}_\perp$, $\theta_\perp$ and $C_\|$, $\chi^{VV}_\|$, $\theta_\|$ for YbRh$_{2}$Si$_{2}$ and YbIr$_{2}$Si$_{2}$ (as shown in Fig. 2 for YbRh$_{2}$Si$_{2}$ $C_\|$ can be neglected within experimental error). As expected, the values of parameters $C_\perp$ and $C_\|$ are smaller than the calculated 
values $C^0_\perp$ and $C^0_\|$. Indeed, as follows from (\ref{eq13}), the renormalization of the susceptibility by the interaction with conduction electrons reduces the value of Curie constants because of the antiferromagnetic sign of the exchange integral $(J^{\|,\perp}<0)$ and, hence, $\lambda_{\|,\perp}<0$. From our fitting the Weiss temperatures are $\theta_\perp=5.43$~K,  $\theta_\|=0.76$~K for YbRh$_{2}$Si$_{2}$ and $\theta_\perp=4.1$~K,  $\theta_\|=1.98$~K for YbIr$_{2}$Si$_{2}$.

\begin{figure}[htbp]
  \centering
 \includegraphics{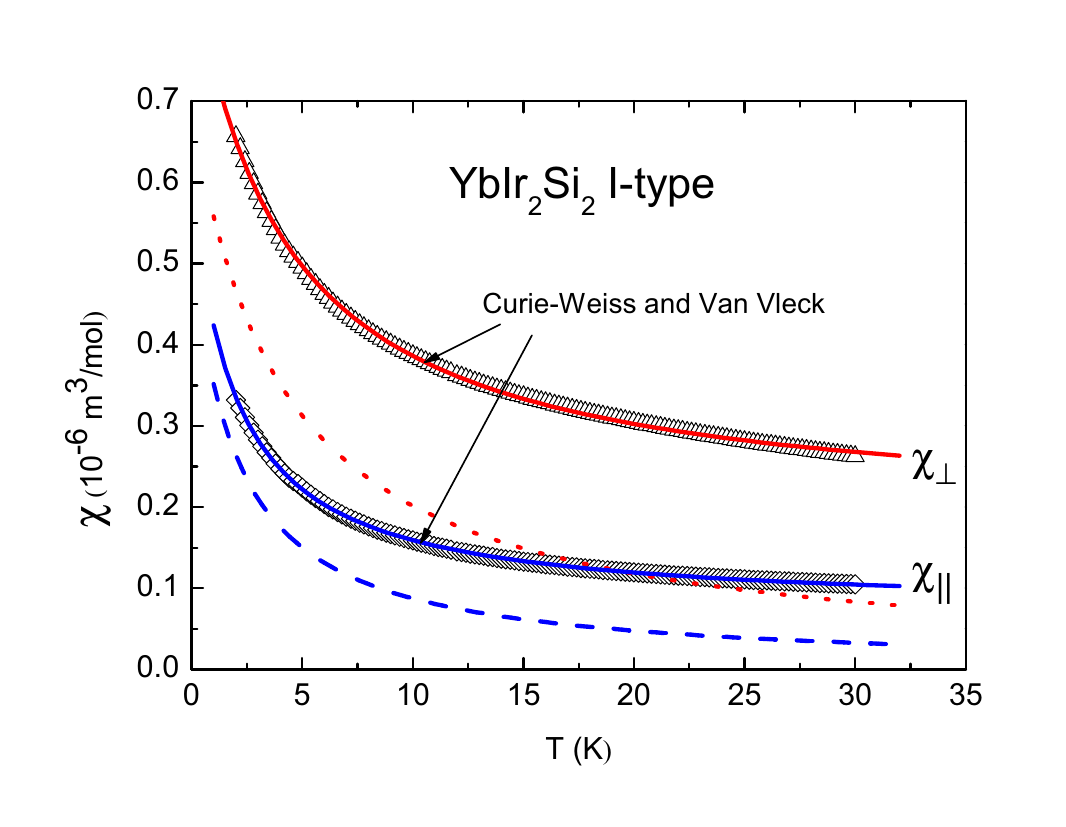}
  \caption{Static magnetic susceptibility of YbIr$_{2}$Si$_{2}$ for $H=10$~kG. Solid lines: fitting of susceptibility data with expression (\ref{eq13}) and parameters shown in Table 4. Dotted and dashed lines: contributions of Curie-Weiss part of susceptibility in perpendicular and in parallel orientations, correspondingly.\label{Fig3}}
 \end{figure}
\Table{\label{tab4}Curie constants $C$ $(10^{-6}\rm{m^3mol^{-1}K})$, Van Vleck susceptibility $\chi^{VV}$ $(10^{-6}\rm{m^3mol^{-1}})$ and Weiss temperatures $\theta(K)$ for YbRh$_{2}$Si$_{2}$ and YbIr$_{2}$Si$_{2}$ from the data fits according Eq.(\ref{eq13}) as shown in Figs. 2-4.}
\br
Compound & $T$ (K) & \centre{1}{$C_\perp$} & \centre{1}{$\chi^{VV}_\perp$} & \centre{1}{$\theta_\perp$} & \centre{1}{$C_\|$} & \centre{1}{$\chi^{VV}_\|$} & \centre{1}{$\theta_\|$} \\
\mr 
                                  & 2--30 & 10.89 & 0.064 & 5.43 & 0.02 & 0.12 & 0.76 \\
\ns \ns
YbRh$_{2}$Si$_{2}$ & & & & & & &\\
\ns \ns
                                  &  0.1--3.6 & 2.31 & 0.75 & 0.22 & \centre{1}{--} & \centre{1}{--} & \centre{1}{--} \\ 
\mr
YbIr$_{2}$Si$_{2}$   & 2--30 &  2.84 & 0.18 & 4.1 & 1.04 & 0.07 & 1.98 \\ \br
\end{tabular}
\end{indented}
\end{table}
\begin{figure}[tbp]
  \centering
 \includegraphics{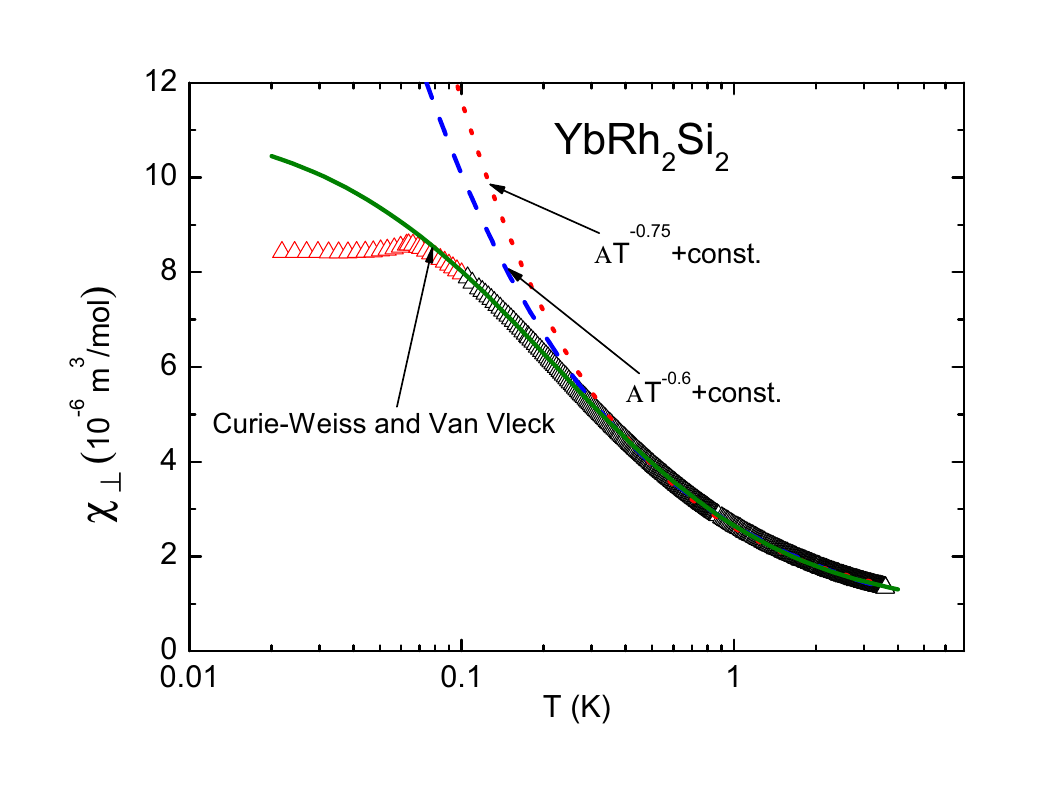}
  \caption{Fitting of the ac susceptibility data without additional dc field from \cite{gegenwart06b} by the power law (\ref{eq15}) with $\alpha= 0.6$ (dotted line), $\alpha=0.75$ (dashed line) and by expression (\ref{eq13}) (solid line) with parameters shown in Table 4.\label{Fig4}}
 \end{figure}
\section{Discussion}
Our calculations of the magnetic susceptibility of YbRh$_{2}$Si$_{2}$ and YbIr$_{2}$Si$_{2}$  on the basis of an entirely local model of the Yb$^{3+}$ ion in the crystal electric field were stimulated by the following reasons. Firstly, the observed EPR signal reflects a number of features, which are very similar to those expected for Yb$^{3+}$ ions doped in non-conducting crystals (in particular, the local crystal field symmetry, the value of $g$ factors, the temperature dependence of the EPR intensity, see \cite{abragam70a}). Even the EPR linewidth shows a temperature dependence that resembles the behaviour of Yb$^{3+}$ ions diluted in a conducting environment \cite{sichelschmidt03a, barnes81a}. Secondly, an intensive experimental study of the NFL magnetic and thermal properties of these materials points out locally critical fluctuations. Thirdly, the calculations for the used local model could be performed in a straightforward and transparent way.\\
Our major result is a remarkable agreement of our local approach for the static magnetic susceptibility with the temperature dependence of the experimental data. Therefore, in the considered region of temperatures ($0.1-30$~K) a ballistic motion of the 4$f$ electrons is practically absent, and they could be considered as quasi-localized. However, when approaching lower temperatures ferromagnetic quantum critical fluctuations dominate \cite{gegenwart08a, gegenwart06b} and a locally quantum critical scenario may be applicable \cite{si01a}. One of the hallmarks of this scenario is a generalized Curie-Weiss law which for a wavevector-dependent magnetic susceptibility can be written in the form
\begin{equation}
\label{eq15}
\chi(\mathbf{q},T)=\frac{C}{T^\alpha+\theta(\mathbf{q})^\alpha}
\end{equation}
with an exponent $\alpha<1$ \cite{si01a,schroder00a}. In the case of YbRh$_{2}$Si$_{2}$ such a behaviour was revealed in the temperature region $0.3<T<10$~K for $\mathbf{q} = 0$, with $\alpha=0.6$ and $\theta=0$ \cite{gegenwart06b}. As shown in Fig. 4, we point out that a Curie-Weiss law together with a Van Vleck contribution convincingly describes the data for a wider temperature region in comparison with equation (\ref{eq15}) and down to temperatures just above the AFM ordering temperature. However, for the temperature region shown in Fig. 4 the fitting parameters for the low temperature region are considerably changed, see Table 4. The reduction of the Curie constant $C$ and Weiss temperature ($\theta=0.22$~K), as well as an increase of the temperature independent contribution can be related to the approach of the system to the LFL regime with a more ballistic motion of the 4$f$ electrons 
and indicate the Kondo effect in the magnetic susceptibility data. In this respect a Curie-Weiss description well within the Kondo regime, i.e. at $T\ll T_K$, despite successful, may appear not appropriate. However, strong ferromagnetic correlations, as indicated, for instance, by a large Sommerfeld-Wilson ratio for YbRh$_{2}$Si$_{2}$ \cite{gegenwart06b} and YbIr$_{2}$Si$_{2}$ \cite{hossain05a}, dominate the magnetic susceptibility and may lead to this Curie-Weiss behavior.
The reduction of the Curie constant can also be observed experimentally when comparing the magnetic susceptibility per Yb ion of YbRh$_{2}$Si$_{2}$ with Y$_{1-x}$Yb$_x$Pd$_3$ ($x=0.6\%$) where the 4$f$ electrons are not hybridized with the conduction electrons \cite{wykhoff07b}. Interestingly, the Yb$^{3+}$ EPR intensity of the YPd$_3$:Yb system compares well with the EPR intensity of YbRh$_{2}$Si$_{2}$ \cite{wykhoff07b}. In respect of this, yet unexplained observation, it is worth to mention that in spite of the success of our entirely local approach for the static magnetic susceptibility of YbRh$_{2}$Si$_{2}$ this model is insufficient for a proper theoretical understanding of the dynamical susceptibilities as observed by EPR. We have found that this problem can be considered by taking into account a translational diffusion of 4$f$ electrons and their collective response together with wide-band conduction electrons to the resonant magnetic alternating field (the bottleneck regime). However, a discussion of this problem is beyond the scope of this paper and results will be published elsewhere.
\ack{This work was supported by the Volkswagen Foundation (I/82203) and partially by the RF President Program ''Leading scientific schools'' 2808.2002.2; A.M.S. was supported partially by cooperative Grant Y4-P-07-15 from the CRDF BRHE Program and from the RF Ministry of Education and Science, RNP. 2.2.2.3.10028.}
\section*{References}
%

\end{document}